\pdfoutput=1
\documentclass[%
 reprint,
 amsmath,amssymb,
 aps,
 nofootinbib,
 superscriptaddress,
 longbibliography,
 showkeys
]{revtex4-2}

\usepackage{amsthm}
\usepackage{graphicx}
\usepackage{xcolor}
\usepackage{braket}
\usepackage{amsmath}
\usepackage{amssymb}
\usepackage{enumerate}
\usepackage{mathtools}
\usepackage{etoolbox}
\usepackage{algorithm}
\usepackage{algpseudocode}
\usepackage{bbm}
\usepackage{float}
\usepackage[
colorlinks=true,
urlcolor=blue,
citecolor=blue,
linkcolor=blue,
hyperfootnotes=false,
breaklinks=true
]{hyperref}
\usepackage{subfigure}

\newtheorem{theorem}{Theorem}

\newtheorem{problem}{Problem}

\begin{document}

\title{On the bias in iterative quantum amplitude estimation}
\author{Koichi Miyamoto}
\email{miyamoto.kouichi.qiqb@osaka-u.ac.jp}
\affiliation{Center for Quantum Information and Quantum Biology, Osaka University, Toyonaka, Osaka 560-0043, Japan}

\date{\today}

\begin{abstract}

Quantum amplitude estimation (QAE) is a pivotal quantum algorithm to estimate the squared amplitude $a$ of the target basis state in a quantum state $\ket{\Phi}$.
Various improvements on the original quantum phase estimation-based QAE have been proposed for resource reduction.
One of such improved versions is iterative quantum amplitude estimation (IQAE), which outputs an estimate $\hat{a}$ of $a$ through the iterated rounds of the measurements on the quantum states like $G^k\ket{\Phi}$, with the number $k$ of operations of the Grover operator $G$ (the Grover number) and the shot number determined adaptively.
This paper investigates the bias in IQAE.
Through the numerical experiments to simulate IQAE, we reveal that the estimate by IQAE is biased and the bias is enhanced for some specific values of $a$.
We see that the termination criterion in IQAE that the estimated accuracy of $\hat{a}$ falls below the threshold is a source of the bias.
Besides, we observe that $k_\mathrm{fin}$, the Grover number in the final round, and $f_\mathrm{fin}$, a quantity affecting the probability distribution of measurement outcomes in the final round, are the key factors to determine the bias, and the bias enhancement for specific values of $a$ is due to the skewed distribution of $(k_\mathrm{fin},f_\mathrm{fin})$.
We also present a bias mitigation method: just re-executing the final round with the Grover number and the shot number fixed.

\end{abstract}


\maketitle

\section{Introduction}

Among various quantum algorithms, {\it quantum amplitude estimation (QAE)} is one of the prominent ones.
It is originally proposed in \cite{brassard2002} as a method to estimate the squared amplitude $a$ of the target basis state $\ket{\phi_1}$ in a quantum state $\ket{\Phi}$.
If we have the oracle $A$ to generate $\ket{\Phi}$ and the reflection operator $S$ with respect to $\ket{\phi_1}$, we can obtain an $\epsilon$-approximation of $a$\footnote{For $x\in\mathbb{R}$, we say that $y\in\mathbb{R}$ is an $\epsilon$-approximation of $x$ if $|y-x|\le\epsilon$.}, querying $A$ and $S$ $O(1/\epsilon)$ times. 
More concretely, the method in \cite{brassard2002} is based on {\it quantum phase estimation (QPE)} \cite{kitaev1995quantum}: using $A$ and $S$, we construct the {\it Grover operator} $G$ (defined later), which acts on $\ket{\Phi}$ to amplify the amplitude of $\ket{\phi_1}$, and then operate the controlled version of $G$ $O(1/\epsilon)$ times followed by an inverse quantum Fourier transform, which yields an approximation of $a$.

One reason why QAE is important is that it is the basis of other quantum algorithms.
For example, it is used in the quantum algorithm for Monte Carlo integration (QMCI) \cite{montanaro2015}, which estimates the expectation of a random variable quadratically faster than the classical counterpart.
Furthermore, QMCI has many applications in industry, e.g., derivative pricing \cite{Rebentrost2018,Stamatopoulos2020optionpricingusing,Chakrabarti2021thresholdquantum,miyamoto2022bermudan,kaneko2022quantum,doriguello2022} 
in finance.

Partly because of such practical importance, many improvements to the original version of QAE have been proposed so far.
In particular, some methods {\it without} QPE have been devised \cite{suzuki2020amplitude,Aaronson2020,nakaji2020faster,grinko2021iterative,tanaka2021amplitude,GiurgicaTiron2022lowdepthalgorithms,Uno_2021,wada2022quantum,Tanaka2022,callison2022improved,Fukuzawa2023,lu2023adaptive}.
Since the controlled $G$ requires a larger gate cost than the uncontrolled one, replacing the former with the latter leads to gate cost reduction.
The first algorithm in such a direction is QAE based on the {\it maximum likelihood estimation (MLE)} \cite{suzuki2020amplitude}, which we hereafter call {\it MLEQAE}.
In this method, we apply $G$ to $\ket{\Phi}$ $k$ times and make $N$ measurements on $G^{k}\ket{\Phi}$ by which we distinguish $\ket{\phi_1}$ and other basis states, increasing $k$ according to some given schedule.
We hereafter call this $k$, the number of operations of $G$, the {\it Grover number}.
The outcomes of the measurements, namely the numbers of times we get $\ket{\phi_1}$ for the various $k$, give us the information on $a_k=\sin^2\left((2k+1)\times \arcsin(\sqrt{a})\right)$, the squared amplitude of $\ket{\phi_1}$ in $G^k\ket{\Phi}$, and thus the information on $a$.
Then, we use this to construct the likelihood function of $a$ and obtain an estimate of $a$ as the maximum likelihood point.

Afterward, the paper \cite{grinko2021iterative} proposed {\it iterative quantum amplitude estimation (IQAE)}, which this paper focuses on.
Like MLEQAE, in IQAE we use the outcomes of the measurements on $G^{k}\ket{\Phi}$ with varying $k$, but in a different way.
Starting from $\ket{\Phi}$, which corresponds to $k=0$, we obtain the confidence interval (CI) of $a$ by the MLE based on the outcomes of the measurements on $\ket{\Phi}$.
We then choose the next $k$ adaptively in the way explained later, and make the measurements on $G^{k}\ket{\Phi}$.
Via MLE, this yields the CI on $a_k$, which is translated into the CI of $a$ narrower than the previous one.
We repeat these steps, each of which is called a {\it round}, increasing the Grover number and narrowing the CI, until the CI width reaches the required accuracy $\epsilon$.
In addition to this adaptive increment of the Grover number, another feature of IQAE is that $N$ the number of the measurements (the {\it shot number}) in one round is increased gradually: if the CI of $a$ derived from measurements with current $k$ is so narrow that we can determine the next $k$, we stop the current round and go to the next round with the next $k$, and otherwise, we add more measurements with the current $k$.
The advantage of such an adaptive increment of $k$ and $N$ is that it can lead to saving the total number of queries to $G$ compared to fixing $k$ and $N$ in advance\footnote{Also in the framework of MLEQAE, some recent studies proposed setting the shot number adaptively \cite{wada2022quantum}.}.

In this paper, we focus on the {\it bias} in IQAE, which can be an issue in some situations but has not been focused on in previous studies.
Note that the estimate $\hat{a}$ of $a$ by IQAE is stochastic since it is derived from outcomes of measurements on quantum states, which are intrinsically random.
Although IQAE guarantees that the magnitude of the error $\hat{a}-a$ is below the tolerance with high probability, it might be biased, that is, the expectation of the error might not be zero: $b:=\mathbb{E}[\hat{a} - a]\ne0$, where $\mathbb{E}[\cdot]$ is the expectation with respect to the randomness of the measurement outcomes.
We hereafter call $b$ the bias and the residual $\hat{a}-a-b$ as the random part.

The motivation to focus on the bias in QAE, including IQAE, is the possibility that it may matter more than random part in some cases.
That is, when we want some quantities given as a combination of many outputs of different QAE runs, biases can accumulate and, even if small in each output, become significant in total.
In worst cases, biases in the sum of $N$ estimates scales as $O(N)$, whereas according to the central limit theorem, the random parts cancel each other and scale as $O(\sqrt{N})$.
An example of situations where we combine many QAE outputs is calculating the total value of a portfolio of derivative contracts, where each contract is priced by individual runs of QAE.
Another example is calculating Gibbs partition functions in statistical mechanics~\cite{Cornelissen2023}.
In the quantum algorithm in \cite{Cornelissen2023}, a partition function is expressed as the product of expectations of certain random variables, each of which is estimated by QAE.


As far as the author knows, there has been no study on bias in IQAE, while bias in other types of QAE has been studied so far \cite{Cornelissen2023,callison2022improved,lu2023adaptive}.
Thus, in this paper, we investigate the bias in IQAE.
We conduct numerical experiments to reveal the nature of the bias in IQAE. 
One of our key findings from the numerical experiments is that IQAE is in fact biased, and its {\it termination criterion} induces the bias.
That is, the procedure that the algorithm ends when the CI width of $a$ reaches the required accuracy leads to the bias.
This is because the CI is statistically inferred and its width depends on the realized value of the estimate $\hat{a}$ of $a$.
The algorithm tends to end when $\hat{a}$ accidentally takes a value that yields a narrow CI.
This effect affects the expectation of $\hat{a}$ and then induces the bias.

In particular, we observe that the bias is enhanced for some specific values of $a$. 
This phenomenon is explained by the distribution of $(k_\mathrm{fin},f_\mathrm{fin})$.
Here, $k_\mathrm{fin}$ is the Grover number in the final round, and $f_\mathrm{fin}\in[0,1]$ is a quantity determined by $k_\mathrm{fin}$ (see the definition later), which rapidly varies by a slight change of $k_\mathrm{fin}$ and largely affects the bias.
Note that $k_\mathrm{fin}$ is also a random variable depending on the measurement outcomes, and thus so is $(k_\mathrm{fin},f_\mathrm{fin})$.
For a value of $a$ other than the specific ones, $f_\mathrm{fin}$ takes the various values distributed widely in the range $[0,1]$ when $k_\mathrm{fin}$ varies.
Then, over the wide distribution of realized values of $(k_\mathrm{fin},f_\mathrm{fin})$ in the 2D plane, the various values of the bias for the various values of $(k_\mathrm{fin},f_\mathrm{fin})$ are canceled out on average, which yields a small bias in total.
On the other hand, for the specific values of $a$, the realized values of $(k_\mathrm{fin},f_\mathrm{fin})$ are not distributed widely but concentrated in a small part of the 2D plane, in fact, on a few curves.
Thus, the bias cancellation does not occur and the resultant bias remains considerable.

We also propose a simple way to mitigate the bias: just re-executing the final round.
If the algorithm ends at the final round with the Grover number $k_\mathrm{fin}$ and the shot number $N_\mathrm{fin}$, we perform another round with the same Grover number and shot number and obtain an estimate of $a$ from the measurement outcomes in this additional round.
Now, the gradual increment of the shot number and the termination criterion on the CI width are no longer adopted: we perform just $N_\mathrm{fin}$ shots and stop.
This largely diminishes the bias by the termination criterion, as confirmed by the numerical experiments.
We also confirm that, although this re-executing solution definitely increases the total number of queries to $G$, the increase rate is modest -- about 25\% in our experiment.
This is because the final round does not dominate the other rounds in terms of the query number.

The rest of this paper is organized as follows.
Section \ref{sec:Prel} is a preliminary one, where we outline QAE and IQAE.
Section \ref{sec:NumExp} is the main part of this paper, where the results of our numerical experiments are presented.
We first show the magnitude of the bias for various values of $a$, which is enhanced for some specific values of $a$, and then elaborate the aforementioned understanding of such a phenomenon.
We finally propose the bias mitigation method by re-executing the final round along with the numerical result.
Section \ref{sec:Sum} summarizes this paper.

\section{Preliminaries \label{sec:Prel}}

\subsection{Quantum amplitude estimation}

In this paper, QAE is a generic term that means quantum algorithms to estimate the amplitude of a target basis state in a superposition state.
Concretely, our aim is described as the following problem.

\begin{problem}
Let $\epsilon,\alpha\in(0,1)$.
Suppose that we are given access to the following two quantum circuits $A$ and $S$ on a $n$-qubit register.
$A$ acts as
\begin{equation}
    A\ket{0} = \ket{\Phi} := \sqrt{a}\ket{\phi_1} + \sqrt{1-a}\ket{\phi_0},
    \label{eq:A}
\end{equation}
where $a\in[0,1]$, $\ket{0}$ is the computational basis state in which all the $n$ qubits take $\ket{0}$, and $\ket{\phi_0}$ and $\ket{\phi_1}$ are quantum states orthogonal to each other.
$S$ acts as
\begin{equation}
    \begin{cases}
    S\ket{\phi_0} & = \ket{\phi_0} \\
    S\ket{\phi_1} & = -\ket{\phi_1}
    \end{cases}.
    \label{eq:R}
\end{equation}
Also suppose that we can measure an observable corresponding to a Hermitian $H$ on the same system, for which $\ket{\phi_0}$ and $\ket{\phi_1}$ are eigenstates with different eigenvalues.
Then, we want to get an estimate of $a$ with accuracy $\epsilon$ with probability at least $1-\alpha$.
\label{prob:qae}
\end{problem}

Although the setup of Problem \ref{prob:qae} seems quite simple, previous studies \cite{brassard2002,suzuki2020amplitude,Aaronson2020,nakaji2020faster,grinko2021iterative,tanaka2021amplitude,GiurgicaTiron2022lowdepthalgorithms,Uno_2021,wada2022quantum,Tanaka2022,callison2022improved,Fukuzawa2023,lu2023adaptive} have generally considered this setup, and in fact, many applications can be boiled down to this form.
For example, in QMCI \cite{montanaro2015}, the expected value of a random variable is encoded into a quantum state like Eq. \eqref{eq:A} as the squared amplitude $a$ of a basis state $\ket{\phi_1}$, and through estimating this amplitude, we get an approximation of the expectation. 

In many use cases of QAE, $\ket{\phi_1}$ and $\ket{\phi_0}$ are distinguished by whether a specific qubit takes $\ket{1}$ or $\ket{0}$.
In this case, $S$ is the $Z$ gate on the qubit and $H$ is the projective measurement in the computational basis on the qubit.

\cite{brassard2002} posed this problem and presented an algorithm for it based on QPE.
We have the following theorem on its query complexity.

\begin{theorem}[\cite{brassard2002}, Theorem 12, modified]
    Suppose that we are given access to the oracles $A$ and $S$ in Eqs. (\ref{eq:A}) and (\ref{eq:R}).
    Then, for any $\epsilon,\alpha\in(0,1)$, there exists a quantum algorithm that outputs $\hat{a}\in(0,1)$ such that $|\hat{a}-a|\le\epsilon$ with probability at least $1-\alpha$ calling $A$ and $S$
    \begin{equation}
        O\left(\frac{1}{\epsilon}\log\left(\frac{1}{\alpha}\right)\right)
    \end{equation}
    times.
    \label{th:QAE}
\end{theorem}

Although the success probability in the original algorithm in \cite{brassard2002} is lower bounded by a constant $8/\pi^2$, it can be enhanced to an arbitrary value $1-\alpha$ at the expense of an $O\left(\log(1/\alpha)\right)$ overhead in the query complexity.
This is done by a trick of taking the median of the results in the multiple runs of the algorithm \cite{montanaro2015}, which is based on the powering lemma in \cite{jerrum1986}.

We do not give the full details of this QPE-based QAE but present its outline briefly.
The key ingredient is the Grover operator $G$, which is defined as
\begin{equation}
    G:=-A S_0 A^\dagger S.
    \label{eq:GrovOp}
\end{equation}
Here, $S_0$ is a unitary such that
\begin{equation}
    S_0 \ket{\phi} = 
    \begin{cases}
    -\ket{0} & ; \ \mathrm{if} \ \ket{\phi}=\ket{0} \\
    \ket{\phi} & ; \ \mathrm{if} \ \braket{\phi | 0} = 0 
    \end{cases},
\end{equation}
which can be implemented as a combination of $X$ gates and a multi-controlled $Z$ gate.
The key property of $G$ is that for any $k\in\mathbb{N}$ it acts as
\begin{equation}
    G^k\ket{\Phi} = \sin\left((2k+1)\theta_a\right)\ket{\phi_1} + \cos\left((2k+1)\theta_a\right)\ket{\phi_0},
    \label{eq:Gk}
\end{equation}
where $\theta_a:=\arcsin\left(\sqrt{a}\right)$.
That is, $G$ rotates the statevector by an angle $2\theta_a$ in the 2-dimensional Hilbert space spanned by $\ket{\phi_1}$ and $\ket{\phi_0}$.
Because of this property, we have the following QPE-based approach to estimate $a$.
We prepare a register $R_G$ on which $G$ acts and an ancillary $m$-qubit register $R_\mathrm{anc}$.
Then, using $G,G^2,\ldots,G^{2^{m-1}}$ controlled by the first, second, ..., $m$-th qubits in $R_\mathrm{anc}$, respectively, we generate the state $\frac{1}{2^{m/2}}\sum_{k=0}^{2^m-1}\ket{k}G^k\ket{\Phi}$.
Finally, we operate the inverse quantum Fourier transform on $R_\mathrm{anc}$, which, thanks to the property in Eq. \eqref{eq:Gk}, yields an $m$-bit precision estimate of $\theta_a$ and thus that of $a=\sin^2 \theta_a$.
In this process, the number of uses of (controlled) $G$ is $1+2+\ldots+2^{m-1}=2^m$, and thus $A$ and $S$ are queried $O(2^m)$ times, which implies the $O(1/\epsilon)$ query complexity for accuracy $\epsilon$ as shown in Theorem \ref{th:QAE}.
See \cite{brassard2002} for more details.

Compared to a naive way for estimating $a$ by repeating generations of $\ket{\Phi}$ and measurements on it and letting the frequency of obtaining $\ket{\phi_1}$ be an estimate of $a$, in which the number of queries to $A$ scales as $\widetilde{O}(1/\epsilon^2)$, QAE achieves the quadratic speedup with respect to $\epsilon$.
This is the origin of the quadratic speedups in quantum algorithms built upon QAE, e.g., QMCI in comparison to the classical Monte Carlo method.

\subsection{Iterative quantum amplitude estimation}

\begin{figure}
\begin{algorithm}[H]
\caption{Modified IQAE}
\label{alg:IQAE}
\begin{algorithmic}[1]
\Require $\epsilon,\alpha \in (0,1)$, $N_{\text{shot}} \in \mathbb{N}$, $r_\mathrm{min}>1$
\State $K_\mathrm{max} \leftarrow \frac{\pi}{4\epsilon}$
\State $k_1\leftarrow0$
\State $\theta^l_\mathrm{last}\leftarrow0$
\For{$i=1,2,...$}
\State $N_{i,0} \leftarrow 0$, $N^1_{i,0} \leftarrow 0$ \label{step:IQAERoundBegin}
\State $K_i \leftarrow 2k_i + 1$
\State $\alpha_i \leftarrow \frac{2\alpha}{3} \frac{K_i}{K_\mathrm{max}}$
\State $N_i^\mathrm{max} \leftarrow \frac{2}{\sin^2(\pi/21)\sin^2(8\pi/21)}\ln\left(\frac{2}{\alpha_i}\right)$
\State
$R_i \leftarrow \lfloor \frac{K_i\theta^l_\mathrm{last}}{\pi/2} \rfloor$
\Statex \Comment{indicates the quadrant enclosing  $[K_i\theta^l_{i,j},K_i\theta^u_{i,j}]$}
\For{$j=1,2,...$}
\State $n_{\mathrm{shot}}\leftarrow\min\{N_{\mathrm{shot}}, N_{i}^\mathrm{max}-N_{i,j-1}\}$ 
\State Iterate generating $G^{k_i}\ket{\Phi}$ and measuring it $n_{\mathrm{shot}}$ times. Let the number of times $\ket{\phi_1}$ is obtained be $n_1$. \label{step:ite}
\State $N_{i,j} \leftarrow N_{i,j-1} + n_{\mathrm{shot}}$, $N^1_{i,j} \leftarrow N^1_{i,j-1} + n_1$
\State $\hat{a}_{k_i,j} \leftarrow \frac{N^1_{i,j}}{N_{i,j}}$
\Statex \Comment{Maximum likelihood estimate of $a_{k_i}$}
\State $\epsilon_{i,j} \leftarrow \sqrt{\frac{1}{2N_{i,j}} \ln\left(\frac{2}{\alpha_i}\right)}$ 
\State
\begin{equation}
    [a^l_{k_i,j}, a^u_{k_i,j}] \leftarrow \left[\max(0, \hat{a}_{k_i,j} - \epsilon_{i,j}),\min(1, \hat{a}_{k_i,j} + \epsilon_{i,j})\right]
    \label{eq:CIofak}
\end{equation}
\Statex \Comment{CI of $a_{k_i}$}
\State \begin{equation}
    \hat{\gamma}_{i,j} \leftarrow  \gamma(\hat{a}_{k_i,j},R_i), 
    \gamma_{i,j}^l \leftarrow \gamma(a^l_{k_i,j},R_i),
    \gamma_{i,j}^u \leftarrow \gamma(a^u_{k_i,j},R_i),
    \label{eq:gamma_a}
\end{equation}
where
\begin{equation}
    \gamma(a^\prime,r):=
    \begin{cases}
         \arcsin\left(\sqrt{a^\prime}\right) & ; \ \mathrm{if} \ r \ \mathrm{is} \ \mathrm{even} \\
        \frac{\pi}{2}-\arcsin\left(\sqrt{a^\prime}\right) & ; \ \mathrm{if} \ r \ \mathrm{is} \ \mathrm{odd}
    \end{cases}
    \label{eq:gamma}
\end{equation}
for $a^\prime\in[0,1]$ and $r\in\mathbb{N}$.

\State $\hat{\theta}_{i,j}\leftarrow\frac{R_i \frac{\pi}{2} + \hat{\gamma}_{i,j}}{K_i}$ 
\Statex \Comment{Maximum likelihood estimate of $\theta_a$}
\State $[\theta^l_{i,j}, \theta^u_{i,j}] \leftarrow \left[\frac{R_i \frac{\pi}{2} + \gamma^l_{i,j}}{K_i}, \frac{R_i \frac{\pi}{2} + \gamma_{i,j}^u}{K_i}\right]$ \Comment{CI of $\theta_a$}
\State $\hat{a}_{i,j}\leftarrow\sin^2\hat{\theta}_{i,j}$ 
\Statex \Comment{Maximum likelihood estimate of $a$}
\State $[a^l_{i,j}, a^u_{i,j}] \leftarrow [\sin^2\theta^l_{i,j}, \sin^2\theta^u_{i,j}]$ \Comment{CI of $a$}
\State 
\begin{equation}
\Delta a_{i,j} \leftarrow \max\{\hat{a}_{i,j}-a^l_{i,j},a^u_{i,j}-\hat{a}_{i,j}\}
\label{eq:Deltaaij}
\end{equation}
\Statex \Comment{Estimated accuracy of $\hat{a}_{i,j}$}
\If{$\Delta a_{i,j} \le \epsilon$} \label{step:IQAETermCri}
\State \Return $\hat{a}_{i,j}$ as $\hat{a}$
\EndIf
\State $k_\mathrm{temp} \leftarrow \texttt{FindNextK}(k_{i}, \theta^l_{i,j}, \theta^u_{i,j}, r_\mathrm{min})$
\If{$k_\mathrm{temp}>k_i$}
\State $k_{i+1}\leftarrow k_\mathrm{temp}$
\State $\theta^l_\mathrm{last} \leftarrow \theta^l_{i,j}$
\State \textbf{break} the inner for-loop 
\EndIf
\EndFor \label{step:IQAERoundEnd}
\EndFor 
\end{algorithmic}
\end{algorithm}
\end{figure}

\begin{figure}[tp]
\begin{algorithm}[H]
\caption{FindNextK}
\label{alg:find-next-k}
\begin{algorithmic}[1]
\Require $k_{i}$, $\theta_l$, $\theta_u$, $r_\mathrm{min}$
\State $K_{i} \leftarrow 2k_{i} + 1$
\State $K \leftarrow \left\lfloor \frac{\pi/2}{\theta_u - \theta_l} \right\rfloor$ 
\If {$K$ is even}
\State $K \leftarrow K - 1$
\EndIf
\While{$K \geq r_\mathrm{min} K_{i}$}
\If {$\left\lfloor \frac{K\theta_l}{\pi/2} \right\rfloor = \left\lceil \frac{K\theta_u}{\pi/2} \right\rceil - 1$}
\State \Return $(K - 1)/2$
\EndIf
\State $K \leftarrow K - 2$
\EndWhile
\State \Return $k_{i}$
\end{algorithmic}
\end{algorithm}
\end{figure}

After the original QAE was proposed, some variants have been proposed so far, aiming at the reduction of the resource.
Avoiding the use of QPE is a common approach since QPE requires the controlled version of $G$, for which the resource for the implementation increases compared to the uncontrolled one.

IQAE \cite{grinko2021iterative} is in such a direction.
The basic idea is as follows.
By iterating the generation of $\ket{\Phi}$ and the measurement on it, we get an estimate of $a$ as the frequency of obtaining $\ket{\phi_1}$ in the measurements.
This is in fact a kind of MLE, since for $\mathrm{Be}(p)$, the Bernoulli distribution with probability of 1 equal to $p$, the maximum likelihood estimate of $p$ from multiple trials is nothing but the realized frequency of 1.
We also have the CI of $a$, which contains the true value of $a$ with high probability.
Next, for some $k\in\mathbb{N}$, we repeat generating $G^k\ket{\Phi}$ and the measurement on it, and from the measurement outcomes we get an estimate of
\begin{equation}
    a_k:=\sin^2\left((2k+1)\theta_a\right)
\end{equation}
and then that of $a$.
Here, due to the periodicity of $a_k$ as the function of $a$, the $2k+1$ different values in $[0,1]$ can be the maximum likelihood estimates of $a$.
However, since we already have the CI from the previous round consisting of the measurements on $\ket{\Phi}$, combining it with the measurements on $G^k\ket{\Phi}$ yields the new CI as a single interval in $[0,1]$, whose width is narrower than the previous one.
We repeat this procedure for a sufficient number of rounds, making the width of the CI narrower.
When the CI width reaches the required accuracy $\epsilon$, we have an estimate of $a$ with an error below $\epsilon$.
The CI width in each round is taken so that the probability that in every round the obtained CI successfully encloses the true value $a$ is at least some required success probability $1-\alpha$.  

Concretely, we present this algorithm as Algorithm \ref{alg:IQAE}.
This is a modification of the algorithm \cite{Fukuzawa2023}, which is already a modified version of the original IQAE algorithm in \cite{grinko2021iterative}.

Leaving the full details of this algorithm to \cite{Fukuzawa2023}, we just present a theorem on the query complexity and some comments.

\begin{theorem}[Theorem 3.1 in \cite{Fukuzawa2023}]
    Let $\epsilon,\alpha\in(0,1)$.
    Suppose that we are given access to the oracles $A$ in Eq. (\ref{eq:A}) and $S$ in Eq. (\ref{eq:R}).
    Then, Algorithm \ref{alg:IQAE} outputs an $\epsilon$-approximation $\hat{a}$ of $a$ with probability at least $1-\alpha$, making $O\left(\frac{1}{\epsilon}\log\frac{1}{\alpha}\right)$ queries to $A$ and $S$ in total.
    \label{th:IQAE}
\end{theorem}

Let us make some comments that help us to understand the outline of the algorithm.
In Algorithm \ref{alg:IQAE}, $[a^l_{k_i,j}, a^u_{k_i,j}]$ and $[\theta^l_{k_i,j}, \theta^u_{k_i,j}]$ are the CIs of $a$ and the angle $\theta_a$, respectively.
In the $i$th round, we set the Grover number to $k_i$, and we calculate the maximum likelihood estimate $\hat{a}_{k_i,j}$ of $a_{k_i}$ and its CI $[a^l_{k_i,j},a^u_{k_i,j}]$ from the outcomes of the measurements on $G^{k_i}\ket{\Phi}$.
They are translated to the maximum likelihood estimate $\hat{a}_{i,j}$ of $a$ and the CIs $[\theta^l_{i,j},\theta^u_{i,j}]$ and $[a^l_{i,j},a^u_{i,j}]$.
The $i$th round ends if we find the Grover number in the next round by the procedure $\texttt{FindNextK}$ shown as Algorithm \ref{alg:find-next-k}.
In this procedure, we search the next $k$ {\it greedily}.
Namely, we set it as large as possible, requiring that $[K\theta^l_{i,j},K\theta^u_{i,j}]$ the CI of $\theta_a$ multiplied by $K=2k+1$ lies in a single quadrant, which enables us to determine the CIs of $a$ and $\theta_a$ as single intervals.
If we cannot find such $k$ in the region that $K \ge r_\mathrm{min} (2k_{i}+1)$, we continue the $i$th round.
When $\Delta a_{i,j}$, the accuracy of the estimate of $a$ from the measurement outcomes so far, goes below the predetermined accuracy level $\epsilon$, the algorithm stops and outputs the estimate at that time.

In the rounds of this process, we operate $G^{k_1},G^{k_2},\ldots$ with $k_i$ increasing exponentially as $K_{i+1} \ge r_{\rm min}K_i$, and its value in the final round is of order $O(1/\epsilon)$.
This means that $G$ is queried $O(1/\epsilon)$ times in total, and so are $A$ and $S$, as stated in Theorem \ref{th:IQAE}.

Note that the parameters $N_\mathrm{shot}$ and $r_\mathrm{min}$ in Algorithm \ref{alg:IQAE} are not mentioned in the statement of Theorem \ref{th:IQAE}.
Although there may be various settings on these that make the algorithm work, we adopt the following setting in this paper.
According to \cite{Fukuzawa2023}, $N_\mathrm{shot}$ can be set to 1, which means that we search the next $k$ every time we make one measurement on $G^{k_i}\ket{\Phi}$, and it reduces the query complexity keeping the accuracy.
We thus set $N_\mathrm{shot}=1$ hereafter.
On $r_\mathrm{min}$, there may be some choices such as 2 in \cite{grinko2021iterative} and 3 in \cite{Fukuzawa2023}, and we adopt the former hereafter.

We also note that a slight modification in Algorithm \ref{alg:IQAE} from the algorithm in \cite{Fukuzawa2023}: the former outputs the maximum likelihood estimate $\hat{a}$ when $\Delta a_{i,j}$ becomes smaller than the required accuracy $\epsilon$, while the latter outputs the midpoint of $[a^l_{i,j},a^u_{i,j}]$ when $\theta^u_{i,j}-\theta^l_{i,j}$, the width of the CI of $\theta_a$, becomes smaller than $2\epsilon$.
The easily checked relationship $|a^u_{i,j}-a^l_{i,j}|\le|\theta^u_{i,j}-\theta^l_{i,j}|$ implies that, if we impose the termination criterion on $\theta_a$ even though we want to guarantee the accuracy of $a$, we may take unnecessarily many iterations.
We thus adopt the termination criterion on $a$, and in fact, we confirmed that this leads to a smaller query number than the criterion on $\theta_a$ in the later numerical experiment, although we will not show the result in the latter setting.

\section{Numerical experiments on the bias in IQAE \label{sec:NumExp}}

\subsection{Biases for various $a$ \label{sec:BiasVara}}

Hereafter, in order to understand the bias in IQAE and how it arises, we conduct some numerical experiments.

First of all, since the current problem is characterized by $a$, the amplitude we want to estimate, we run IQAE and see the bias for the various values of $a$.
Here, ``run" does not mean running Algorithm \ref{alg:IQAE} on a real quantum computer or a quantum circuit simulator but a classical simulation of the algorithm based on the probability distribution of the outcomes of the measurements made in the algorithm.
Concretely, we replace the step \ref{step:ite} in Algorithm \ref{alg:IQAE} with ``Draw $n_\mathrm{shot}$ samples from $\mathrm{Be}\left(a_{k_i}\right)$ and let the number of 1's be $n_1$".
Note that, if we know the value of $a$, we know the probability distribution of the outcome in measuring $G^{k_i}\ket{\Phi}$, that is, $\ket{\phi_1}$ with probability $a_{k_i}$ and $\ket{\phi_0}$ with probability $1-a_{k_i}$, which is equivalent to $\mathrm{Be}\left(a_{k_i}\right)$.
Therefore, with the above replacement, we can produce the output of Algorithm \ref{alg:IQAE} under the probability distribution it obeys.
We hereafter call this simulation procedure Algorithm \ref{alg:IQAE}$^\prime$.

\begin{figure}[tp]
\centering
\includegraphics[width=1\columnwidth]{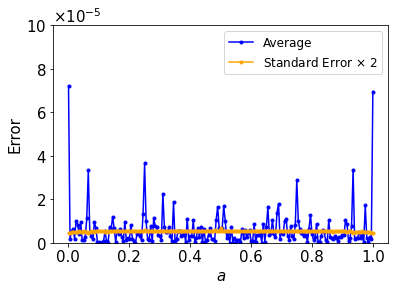}
\caption{For the various values of the amplitude $a$, $|\bar{b}(a)|$, the absolute value of the average of the errors in the 10,000 runs of IQAE (Eq. (\ref{eq:avgErr})) is plotted in blue, and $2\bar{\sigma}(a)$, the standard error of this average (Eq. (\ref{eq:StDevErr})) times $2$ is plotted in orange. We set $\epsilon=0.001,\alpha=0.05$.}
\label{fig:BiasForVariousA}
\end{figure}

For the estimator $\hat{a}$ of $a$, we define its bias as
\begin{equation}
b(a):=\mathbb{E}[\hat{a}-a],    
\end{equation}
where $\mathbb{E}[\cdot]$ denotes the expectation with respect to the randomness of the measurement outcomes in Algorithm \ref{alg:IQAE}.
In order to estimate the magnitude of the bias for various values of $a$, we ran Algorithm \ref{alg:IQAE}$^\prime$ $N_\mathrm{run}$ times, and calculate the value of the average of the errors in the runs
\begin{equation}
    \bar{b}(a):=\frac{1}{N_\mathrm{run}}\sum_{n=1}^{N_\mathrm{run}} \left(\hat{a}^{(n)}-a\right)
    \label{eq:avgErr}
\end{equation}
and the standard error of $\bar{b}(a)$
\begin{equation}
    \bar{\sigma}(a):=\frac{1}{\sqrt{N_\mathrm{run}}}\left[\frac{1}{N_\mathrm{run}}\sum_{n=1}^{N_\mathrm{run}} \left(\hat{a}^{(n)}-a\right)^2\right]^{1/2},
    \label{eq:StDevErr}
\end{equation}
where $\hat{a}^{(n)}$ is the output in the $n$th run.
We can grasp the magnitude of the bias by comparing $\bar{b}(a)$ and $\bar{\sigma}(a)$, since, if $b(a)=0$, $\bar{b}(a)$ is comparable to $\bar{\sigma}(a)$ with high probability, e.g., $|\bar{b}(a)|\le2\bar{\sigma}(a)$ with about 95{\%}, for sufficiently large $N_\mathrm{run}$.
The result is shown in Figure \ref{fig:BiasForVariousA}.
In this figure, we set $a$ to the 201 equally spaced points in $[0.001,0.999]$ including the ends.
The other parameters are set as $N_\mathrm{run}=10^4,\epsilon=0.001,\alpha=0.05$, which also applies hereafter.
From this figure, we see that, for a considerable fraction of the examined values of $a$, $|\bar{b}(a)|$ exceeds $2\bar{\sigma}(a)$, which implies $\hat{a}$ is biased.
In particular, $\bar{b}(a)$ takes much larger values for specific values of $a$ than other values.
Namely, the bias is enhanced for some specific values of $a$.

\subsection{Reason why the bias is enhanced for specific values of $a$}

We then investigate the reason why the bias is enhanced for specific values of $a$, fixing $a$ to $0.2505$, one of such values.
What makes the situation that $a=0.2505$ different from the others?

To make the investigation as simple as possible, we should note some points.
First, we note that it is sufficient to focus on the final round in Algorithm \ref{alg:IQAE}.
The output of Algorithm \ref{alg:IQAE} is determined by the result of the MLE in the final round.
Therefore, as long as the CI obtained in the round just before the final one successfully encloses $a$, which occurs with high probability at least $1-\alpha$, the error is determined by the final round only.
Second, we note that the final round is characterized by only $k_\mathrm{fin}$, the Grover number in that round: other quantities that define the procedure in the final round are automatically determined when $k_\mathrm{fin}$ is fixed, including $R_\mathrm{fin}$, as long as the CI of $a$ encloses its true value.
We finally make a note on notation: here and hereafter, like $k_\mathrm{fin}$ and $R_\mathrm{fin}$, when we write quantities that have $i_\mathrm{fin}$, the index indicating the final round, in the subscript, we replace $i_\mathrm{fin}$ with $\mathrm{fin}$ for conciseness.

Based on the above discussion and noting that $k_\mathrm{fin}$ is also a random variable affected by the measurement outcomes in the previous rounds, we write the bias as
\begin{equation}
    b(a)=\mathbb{E}\left[\hat{a}-a\right]=\sum_{k_\mathrm{fin}^\prime=0}^\infty p_{k_\mathrm{fin}^\prime} b_{k_\mathrm{fin}^\prime}(a)
    \label{eq:BiasDecomp}
\end{equation}
\begin{equation}
    b_{k_\mathrm{fin}^\prime}(a):=\mathbb{E}\left[\hat{a}-a | \mathcal{G}_{k_\mathrm{fin}^\prime}\right],
\end{equation}
where $\mathcal{G}_{k_\mathrm{fin}^\prime}$ is the event that $k_{\mathrm{fin}}$ takes a value $k_\mathrm{fin}^\prime$, and $p_{k_\mathrm{fin}^\prime}$ is its probability.

\subsubsection{Bias conditioned on the final Grover number}

We then consider how $k_\mathrm{fin}$ affects $b_{k_\mathrm{fin}}(a)$, the bias conditioned on $k_\mathrm{fin}$.
First, we rewrite $\hat{a}$ the output of Algorithm \ref{alg:IQAE} as
\begin{equation}
    \hat{a}=\sin^2\left(\frac{R_\mathrm{fin} \frac{\pi}{2} + \hat{\gamma}_\mathrm{fin}}{K_\mathrm{fin}}\right),
    \label{eq:hata_of_gamma}
\end{equation}
where $\hat{\gamma}_\mathrm{fin}$ is the value of $\hat{\gamma}_{\mathrm{fin},j}$ when
\begin{align}
    &a^u_{\mathrm{fin},j}-\hat{a}_{\mathrm{fin},j} \nonumber \\
    &=\sin^2\left(\frac{R_\mathrm{fin} \frac{\pi}{2} + \gamma^u_{\mathrm{fin},j}}{K_\mathrm{fin}}\right)-\sin^2\left(\frac{R_\mathrm{fin} \frac{\pi}{2} + \hat{\gamma}_{\mathrm{fin},j}}{K_\mathrm{fin}}\right)\le \epsilon, \nonumber \\
    &\hat{a}_{\mathrm{fin},j}-a^l_{\mathrm{fin},j} \nonumber \\
    &=\sin^2\left(\frac{R_\mathrm{fin} \frac{\pi}{2} + \hat{\gamma}^l_{\mathrm{fin},j}}{K_\mathrm{fin}}\right)-\sin^2\left(\frac{R_\mathrm{fin} \frac{\pi}{2} + \gamma^l_{\mathrm{fin},j}}{K_\mathrm{fin}}\right)\le \epsilon
    \label{eq:endcond_gamma}
\end{align}
holds for the first time.
Note that $\gamma^l_{\mathrm{fin},j},\gamma^u_{\mathrm{fin},j}$ and $\hat{\gamma}_{\mathrm{fin},j}$ are random variables determined by the measurement outcomes in Algorithm \ref{alg:IQAE}, and so is $\hat{\gamma}_\mathrm{fin}$.
When $k_\mathrm{fin}\gg1$, which is a typical situation for small $\epsilon$, the most sensitive dependence of $\hat{a}$'s distribution on $k_\mathrm{fin}$ is through the distributions of $\gamma^l_{\mathrm{fin},j},\gamma^u_{\mathrm{fin},j}$ and $\hat{\gamma}_{\mathrm{fin},j}$, while $k_\mathrm{fin}$ also affects $\hat{a}$ through $K_\mathrm{fin}$ and $R_\mathrm{fin}$ in Eqs. (\ref{eq:hata_of_gamma}) and (\ref{eq:endcond_gamma}).
This is because the distributions of $\hat{\gamma}_{\mathrm{fin},j}$ etc. largely change even when $k_\mathrm{fin}$ changes slightly. 
Under the definition in Eq. (\ref{eq:gamma_a}), the distributions of $\hat{\gamma}_{\mathrm{fin},j}$ etc. are determined by the distribution of $\hat{a}_{k_\mathrm{fin},j}$ and $\mathfrak{p}(R_\mathrm{fin})$, the parity of $R_\mathrm{fin}$.
The distribution of $\hat{a}_{k_\mathrm{fin},j}$ is determined by that of the outcome of the measurement on $G^{k_\mathrm{fin}}\ket{\Phi}$, which is equivalent to $\mathrm{Be}(a_{k_\mathrm{fin}})$, and $a_{k_\mathrm{fin}}$ can change largely even by shifting $k_\mathrm{fin}$ by 1.
When $\mathfrak{p}(R_\mathrm{fin})$ flips, the form of $\hat{\gamma}_{\mathrm{fin},j}$ as a function of $\hat{a}_{k_\mathrm{fin},j}$ also flips and so does the distribution of $\hat{\gamma}_{\mathrm{fin},j}$.
These observations imply that we should focus on $a_{k_\mathrm{fin}}$ and $\mathfrak{p}(R_\mathrm{fin})$ as key factors for $b_{k_\mathrm{fin}}(a)$.
However, as an equivalent to this, we instead focus on how $b_{k_\mathrm{fin}}(a)$ is affected by
\begin{equation}
    f_\mathrm{fin}:=\mathrm{frac}\left(\frac{(2k_\mathrm{fin}+1)\theta_a}{\pi}\right),
    \label{eq:f_fin}
\end{equation}
where $\mathrm{frac}(x):=x-\lfloor x \rfloor$ is the fractional part of $x\in\mathbb{R}$.
$f_\mathrm{fin}$ has a one-to-one correspondence to the pair $(a_{k_\mathrm{fin}},\mathfrak{p}(R_\mathrm{fin}))$, and taking a single quantity $f_\mathrm{fin}$ as a key factor rather than the pair $(a_{k_\mathrm{fin}},\mathfrak{p}(R_\mathrm{fin}))$ makes the following discussion simpler.

\begin{figure}[tp]
\begin{algorithm}[H]
\caption{Estimate the bias with $k_\mathrm{fin}$ and $f_\mathrm{fin}$ fixed}
\label{alg:EstimBias}
\begin{algorithmic}[1]
\Require $k_\mathrm{fin}\in\mathbb{N}$, $f_\mathrm{fin}\in[0,1]$, $a\in(0,1)$, $N_\mathrm{run}\in\mathbb{N}$, $\epsilon,\alpha \in (0,1)$, $N_{\text{shot}} \in \mathbb{N}$, $r_\mathrm{min}>1$
\State
\begin{align}
    \tilde{R} &\leftarrow \left\lfloor \frac{(2k_\mathrm{fin}+1)\theta_a}{\pi}\right\rfloor \nonumber \\
    \tilde{a}&\leftarrow \sin^2\left(\frac{\left(\tilde{R}+f_\mathrm{fin}\right)\pi}{2k_\mathrm{fin}+1}\right)
\end{align}
\For{$c=1,...,N_\mathrm{run}$}
\State Run one round of IQAE (Steps \ref{step:IQAERoundBegin}-\ref{step:IQAERoundEnd} in Algorithm \ref{alg:IQAE}$^\prime$), setting $k_{i}\leftarrow k_\mathrm{fin}$ and
\begin{equation}
    R_i\leftarrow\left\lfloor \frac{(2k_\mathrm{fin}+1)\theta_{\tilde{a}}}{\pi/2}\right\rfloor,
    \label{eq:RiOnlyFinal}
\end{equation}
and adjusting $a$ to $\tilde{a}$.
\If{the round ends with $\Delta a_{i,j}\le\epsilon$}
\State Let $\tilde{a}_c$ be the value of $\hat{a}_{i,j}$ at the end of the round.
\State $I^\mathrm{end}_c\leftarrow1$
\Else
\State $\tilde{a}_c\leftarrow \textrm{NaN (not a number)}$
\State $I^\mathrm{end}_c\leftarrow0$
\EndIf
\EndFor
\State $N_\mathrm{end}\leftarrow\sum_{c=1}^{N_\mathrm{run}}I^\mathrm{end}_c$
\State Output
\begin{align}
    &\tilde{b}(k_\mathrm{fin}, f_\mathrm{fin}) \nonumber \\
    & \quad \leftarrow
    \begin{dcases}
        \frac{1}{N_\mathrm{end}}\sum_{\substack{c=1,...,N_\mathrm{run} \\ I^\mathrm{end}_c=1}}(\tilde{a}_c-\tilde{a}) & ; \ \mathrm{if} \ N_\mathrm{end}\ge0.1 \times N_\mathrm{run} \\
        \textrm{NaN} & ; \ \textrm{otherwise}
    \end{dcases}.
\end{align}
\end{algorithmic}
\end{algorithm}
\end{figure}

Since $f_\mathrm{fin}$ is also a quantity rapidly changing with respect to $k_\mathrm{fin}$, in order to understand how it affect $b_{k_\mathrm{fin}}(a)$, we temporarily deal with $k_\mathrm{fin}$ and $f_\mathrm{fin}$ as independent variables, even though $k_\mathrm{fin}$ determines $f_\mathrm{fin}$.
We set $k_\mathrm{fin}$ and $f_\mathrm{fin}$ separately, 
and, as an estimate of $b_{k_\mathrm{fin}}(a)$, calculate the quantity $\tilde{b}(k_\mathrm{fin}, f_\mathrm{fin})$ by the procedure in Algorithm \ref{alg:EstimBias}.
That is, we run one round of IQAE many times for fixed $k_\mathrm{fin}$ and $f_\mathrm{fin}$, and take the average of the resultant errors in the runs in which IQAE itself is terminated, neglecting the other runs, which move to the next round.
We calculate the average only when at least 1,000 runs out of the 10,000 total runs yield outputs other than NaN, since averaging a small number of random results leads to an inaccurate estimate.
Note that, in compensation for setting $k_\mathrm{fin}$ and $f_\mathrm{fin}$ independently, the true amplitude $a$ (now, 0.2505) is slightly adjusted to $\tilde{a}$ so that Eq. (\ref{eq:f_fin}) holds.
Nevertheless, we expect that $\tilde{b}(k_\mathrm{fin}, f_\mathrm{fin})$ gives us a good illustration of the behavior of $b_{k_\mathrm{fin}}(a)$.
Also note that the setting in Eq. (\ref{eq:RiOnlyFinal}) corresponds to the assumption that the true amplitude $\tilde{a}$ is enclosed in the CI in the previous round.

\begin{figure*}[tp]
\centering
    \subfigure[$k_\mathrm{fin}=200$]{
    \includegraphics[width=1\columnwidth]{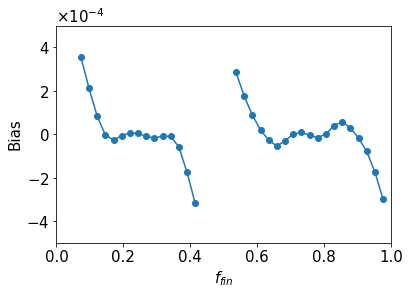} 	\label{fig:bias_kfin200}
    }
    \subfigure[$k_\mathrm{fin}=300$]{
    \includegraphics[width=1\columnwidth]{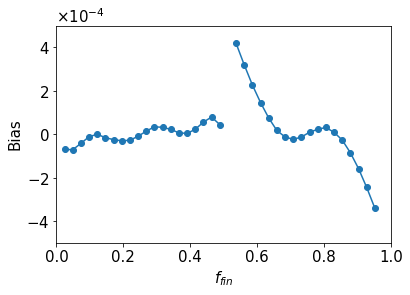} 	\label{fig:bias_kfin300}
    }
    \subfigure[$k_\mathrm{fin}=400$]{
    \includegraphics[width=1\columnwidth]{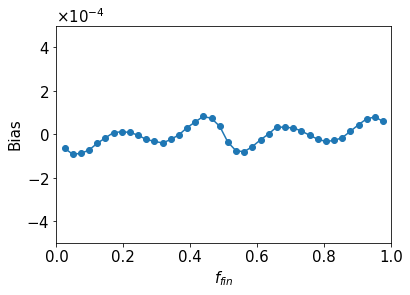} 	\label{fig:bias_kfin400}
    }
    \subfigure[$k_\mathrm{fin}=500$]{
    \includegraphics[width=1\columnwidth]{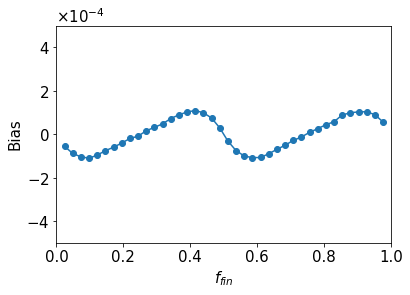} 	\label{fig:bias_kfin500}
    }

\caption{$\tilde{b}(k_\mathrm{fin}, f_\mathrm{fin})$ calculated by Algorithm \ref{alg:EstimBias} for various values of $(k_\mathrm{fin}, f_\mathrm{fin})$. $\tilde{b}(k_\mathrm{fin}, f_\mathrm{fin})$ indicates the bias generated in the final round of IQAE conditioned on the Grover number $k_\mathrm{fin}$ and $f_\mathrm{fin}$ in Eq. \eqref{eq:f_fin}. For $(k_\mathrm{fin},f_\mathrm{fin})$ that yields NaN $\tilde{b}(k_\mathrm{fin}, f_\mathrm{fin})$, we do not plot a point.}
\label{fig:BiasSim}
\end{figure*}

In Figure \ref{fig:BiasSim}, we show the results for the various values of $f_\mathrm{fin}$ with $k_\mathrm{fin}$ set to several values.
For $(k_\mathrm{fin},f_\mathrm{fin})$ that yields NaN $\tilde{b}(k_\mathrm{fin}, f_\mathrm{fin})$, we do not plot a point.

We can intuitively understand how the bias is generated by noting the following two points.

\begin{figure*}[tp]
\centering
    \subfigure[$a^u_{i,j}$ and $a^l_{i,j}$, upper and lower ends of the CI.]{
    \includegraphics[width=1\columnwidth]{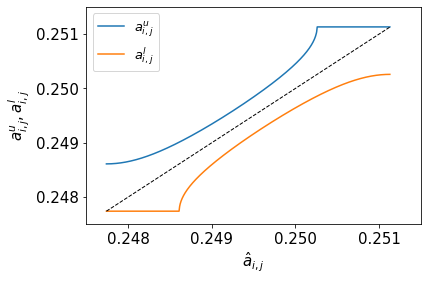} 	\label{fig:CIUBLB}
    }
    \subfigure[Estimated accuracy $\Delta a_{i,j}$.]{
    \includegraphics[width=1\columnwidth]{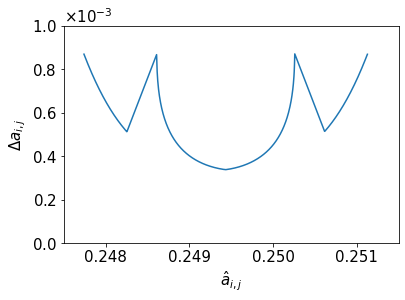} 	\label{fig:CIWidth}
    }

\caption{In Figure \ref{fig:CIUBLB}, the blue (resp. red) solid line is the upper end $a^u_{i,j}$ (resp. lower end $a^l_{i,j}$) of the CI of the amplitude $a$ versus the realized value of the estimate $\hat{a}_{i,j}$ after $N=100$ shots in an IQAE round with the Grover number $k=200$. The black dashed line is a diagonal line just for reference, on which the vertical coordinate is equal to $\hat{a}_{i,j}$. Figure \ref{fig:CIWidth} shows the estimated accuracy $\Delta a_{i,j}$, which is determined by $a^u_{i,j}$, $a^l_{i,j}$, and $\hat{a}_{i,j}$ as Eq. \eqref{eq:Deltaaij}, in the same setting. In these figures, the curves are shown over the range of $\hat{a}_{i,j}$ such that $\left\lfloor (2k+1)\theta_{\hat{a}_{i,j}}/(\pi/2)\right\rfloor=\left\lfloor (2k+1)\theta_{a=0.2505}/(\pi/2)\right\rfloor$. This is the set of the values $\hat{a}_{i,j}$ can take in the IQAE round if the CI at the beginning of the round encloses $a=0.2505$.}
\label{fig:CIUBLBWidth}
\end{figure*}

First, we note that $\Delta a_{i,j}$, the estimated accuracy in the intermediate step in Algorithm \ref{alg:IQAE}, is also a random variable.
It is determined by $\hat{a}_{k_i,j}$, the realized frequency of $\ket{\phi_1}$ in the repeated measurements on $G^{k_i}\ket{\Phi}$, and $N_{i,j}$, the number of the measurements.
How $\Delta a_{i,j}$ depends on $\hat{a}_{k_i,j}$ is understood as follows. 
Because of the upper bound 1 and lower bound 0 of $\hat{a}_{k_i,j}$, the width of the CI $[a^l_{k_i,j},a^u_{k_i,j}]$ of $a_{k_i}$ given as Eq. (\ref{eq:CIofak}) already depends on not only $N_{i,j}$ but also $\hat{a}_{k_i,j}$: if $\hat{a}_{k_i,j}$ is closer to $0$ or $1$, the CI width becomes smaller.
In addition to this, the derivation of $\Delta a_{i,j}$ from $\hat{a}_{k_i,j}$ and $[a^l_{k_i,j},a^u_{k_i,j}]$ by the nonlinear relationship introduce the dependence of $\Delta a_{i,j}$ on $\hat{a}_{k_i,j}$.
Besides, since $\hat{a}_{k_i,j}$ has the one-to-one correspondence to $\hat{a}_{i,j}$, we can regard $\Delta a_{i,j}$ as a function of $\hat{a}_{i,j}$. 
Figure \ref{fig:CIUBLBWidth} illustrates this.
Figure \ref{fig:CIUBLB} shows $a^u_{i,j}$ and $a^l_{i,j}$, the upper and lower ends of the CI of $a$ versus the realized value of $\hat{a}_{i,j}$ after $N=100$ shots in a round with the Grover number $k=200$, and Figure \ref{fig:CIWidth} shows $\Delta a_{i,j}$ versus $\hat{a}_{i,j}$.

The second point is the termination criterion $\Delta a_{i,j}\le \epsilon$ in Step \ref{step:IQAETermCri} in Algorithm \ref{alg:IQAE}, which means that IQAE ends when the estimated accuracy $\Delta a_{i,j}$ becomes smaller than $\epsilon$.
This criterion, along with the aforementioned dependence of $\Delta a_{i,j}$ on $\hat{a}_{i,j}$, leads to the bias.
If $\hat{a}_{i,j}$ goes to the region where $\Delta a_{i,j}$ is relatively small, Algorithm \ref{alg:IQAE} tends to end early, and, if $\hat{a}_{i,j}$ goes to the region with larger $\Delta a_{i,j}$, the algorithm tends to take more shots by the end, or even go to the next round.
This makes the probability that the algorithm ends with $\hat{a}_{i,j}$ corresponding to smaller $\Delta a_{i,j}$ and outputs such $\hat{a}_{i,j}$ as $\hat{a}$ higher.
Such an effect yields the probability distribution of $\hat{a}$ asymmetric about the point $\hat{a}=a$, and the nonzero bias $b_{k_\mathrm{fin}}\ne0$.

Although this understanding on how the bias arises is an intuitive one, we will indirectly confirm its validity by seeing in the numerical experiment in Sec. \ref{sec:miti} that re-executing the final round without the termination criterion mitigates the bias.

With the above understanding, we can also see why the bias $\tilde{b}(k_\mathrm{fin}, f_\mathrm{fin})$ depends on $f_\mathrm{fin}$ with $k_\mathrm{fin}$ fixed.
As explained above, changing $f_\mathrm{fin}$ corresponds to changing $a$.
Assuming that $\hat{a}$ mainly distributes in the neighborhood of $a$, the shape of $\Delta a_{i,j}$ as a function of $\hat{a}_{i,j}$ around the point $\hat{a}_{i,j}=a$ affects the bias, which causes the dependence of the bias on $a$, and then $f_\mathrm{fin}$.
For example, if $\Delta a_{i,j}$ is increasing in the neighborhood of $\hat{a}_{i,j}=a$, the preference of $\hat{a}_{i,j}$ to values corresponding to small $\Delta a_{i,j}$ leads to the negative bias.

We also note that, in Figure \ref{fig:BiasSim}, the graph of $\tilde{b}(k_\mathrm{fin}, f_\mathrm{fin})$ is nearly antisymmetric with respect to reflection about the vertical line $f_\mathrm{fin}=\frac{1}{2}$, that is, 
\begin{equation}
\tilde{b}(k_\mathrm{fin}, f_\mathrm{fin})\approx-\tilde{b}(k_\mathrm{fin}, 1-f_\mathrm{fin}),
\label{eq:biasAntisym}
\end{equation}
except the case of $k_\mathrm{fin}=300$.
This antisymmetricity is understood as follows.
We temporarily ignore the process for the transition to the next round. 
Then, note that, with $k_\mathrm{fin}$ fixed, the transform $f_\mathrm{fin}\rightarrow1-f_\mathrm{fin}$ conserves $a_{k_\mathrm{fin}}$ and flips $\mathfrak{p}(R_\mathrm{fin})$.
Thus, by this transform, the distribution of $\hat{a}_{k_\mathrm{fin},j}$ is unchanged and the relationship between $\hat{\gamma}_{\mathrm{fin},j}$ and $\hat{a}_{k_\mathrm{fin},j}$ in Eq. (\ref{eq:gamma}) is switched.
Therefore, assuming that $k_\mathrm{fin}\gg1$ and neglecting the slight change of $a$ by this transform, we see that this transform flips the distribution of the error $\hat{a}-a$: $p(\hat{a}-a;k_\mathrm{fin},f_\mathrm{fin}) \approx p(-(\hat{a}-a);k_\mathrm{fin},1-f_\mathrm{fin})$, where $p(\cdot;k_\mathrm{fin},f_\mathrm{fin})$ is the probability density of $\hat{a}-a$ conditioned by $k_\mathrm{fin}$ and $f_\mathrm{fin}$.
This means that the transform $f_\mathrm{fin}\rightarrow1-f_\mathrm{fin}$ also flips the bias as Eq. (\ref{eq:biasAntisym}).
In fact, this discussion is not rigorous, since the process for the transition to the next round breaks the symmetry with respect to $f_\mathrm{fin}\rightarrow1-f_\mathrm{fin}$.
This makes the antisymmetricity disappear in Figure \ref{fig:bias_kfin300} for $k_\mathrm{fin}=300$.
Nevertheless, this antisymmetricity holds for the wide region of $(k_\mathrm{fin},f_\mathrm{fin})$, and becomes a key for the phenomenon that the bias is enhanced only for the specific values of $a$, as seen below.

\subsubsection{Distribution of $(k_\mathrm{fin},f_\mathrm{fin})$}

\begin{figure*}[tp]
\centering
    \subfigure[$a=0.2505$.]{
    \includegraphics[width=1\columnwidth]{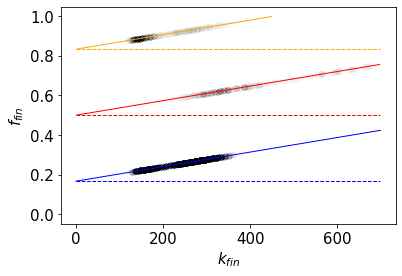} 	\label{fig:kFin_fFin_p02505}
    }
    \subfigure[$a=0.2006$.]{
    \includegraphics[width=1\columnwidth]{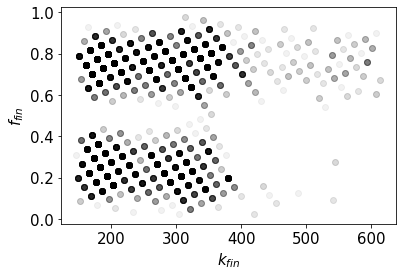} 	\label{fig:kFin_fFin_p02006}
    }

\caption{The realized values of the Grover number $k_\mathrm{fin}$ in the final round and $f_\mathrm{fin}$ given by Eq. \eqref{eq:f_fin} in the 10,000 runs of IQAE for (a) $a=0.2505$ and (b) $a=0.2006$ are plotted as transparent black points. Darker regions consisting of many overlapped points indicate that there are many realizations in it and lighter regions indicates the opposite. In Figure \ref{fig:kFin_fFin_p02505}, the blue, red, and orange solid lines represent $\mathrm{frac}\left((2k_{\rm fin}+1)\theta_a/\pi\right)$ for $k_{\rm fin}$ such that $k_{\rm fin}\equiv0,1$, and $2$ modulo 3, respectively, and the blue, red, and orange horizontal dashed lines represent $f_\mathrm{fin}=1/6$, $1/2$, and $5/6$, respectively.}
\label{fig:ffin}
\end{figure*}

Next, we consider the distribution of $(k_\mathrm{fin},f_\mathrm{fin})$.
Since $f_\mathrm{fin}$ is determined by $k_\mathrm{fin}$ with $a$ fixed and $k_\mathrm{fin}$ takes natural numbers, $(k_\mathrm{fin},f_\mathrm{fin})$ distributes in the 2D plane not continuously but as discrete points.
In Figure \ref{fig:kFin_fFin_p02505}, we show the realized values of $(k_\mathrm{fin},f_\mathrm{fin})$ for $a=0.2505$ in the 10,000 runs of Algorithm \ref{alg:IQAE}$^\prime$.
We see that the plotted points are concentrated only on the three lines.
This is not a phenomenon observed for general $a$.
For example, in Figure \ref{fig:kFin_fFin_p02006}, the similar plot for $a=0.2006$, for which the observed bias $\bar{b}(a)$ is much smaller than $a=0.2505$, the points are distributed more broadly than those for $a=0.2505$.

The distribution of the points for $a=0.2505$ is understood as follows.
We note that, for $\theta\in(0,\pi/2)$ written as $\theta=l\pi/2m$ with positive integers $l$ and $m$ such that $l < m$, $f_\theta(k):=\mathrm{frac}\left((2k+1)\theta/\pi\right)$ is a periodic function of $k\in\mathbb{Z}$ with period $m$.
Thus, it takes at most $m$ values, and, if we plot $f_\theta(k)$ versus $k$, the points lie on at most $m$ horizontal lines.
If we slightly shift $\theta$ from such a value, $f_\theta(k)$ can take different values from the original ones, which means the horizontal lines transform into lines with small slopes.
$\theta_a$ for $a=0.2505$ applies to such a case.
It can be written in the form of
\begin{equation}
    \theta_{a}=\frac{l\pi}{2m}+\delta
    \label{eq:LargeBiasTheta}
\end{equation}
with small $m\in\mathbb{N}$ and $\delta\in\mathbb{R}$ such that $|\delta|\ll1$.
Concretely, $l=1$, $m=3$ and $\delta\approx0.000577$.
For this $a$, $f_{\theta_a}(k)$ consists of only three slightly tilted lines shown in Figure \ref{fig:kFin_fFin_p02505} as solid lines, which were dashed horizontal lines in the same figure if $\delta=0$.

\begin{figure*}[tp]
\centering
    \subfigure[$a=0.2505$.]{
    \includegraphics[scale=0.45]{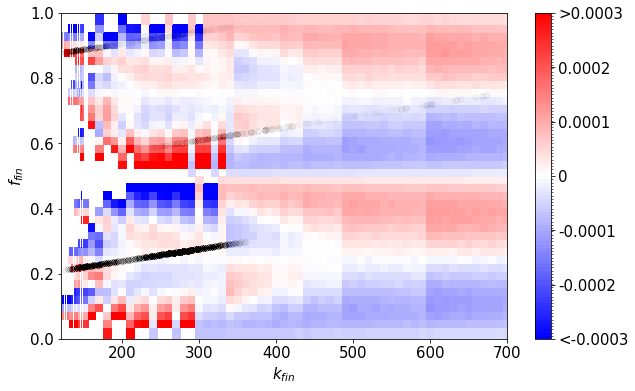} 	\label{fig:heatmap_p02505}
    }
    \subfigure[$a=0.2006$.]{
    \includegraphics[scale=0.45]{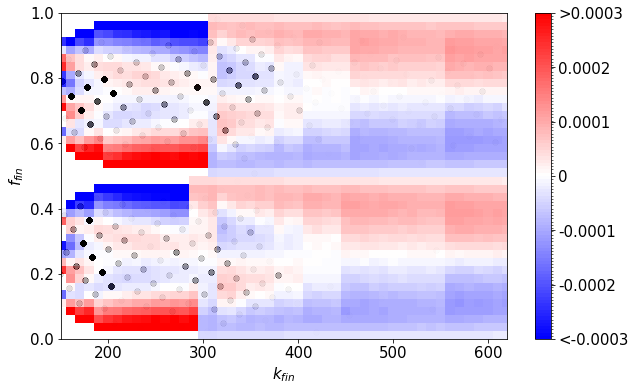} 	\label{fig:heatmap_p02006}
    }
        \subfigure[$a=0.25$.]{
    \includegraphics[scale=0.45]{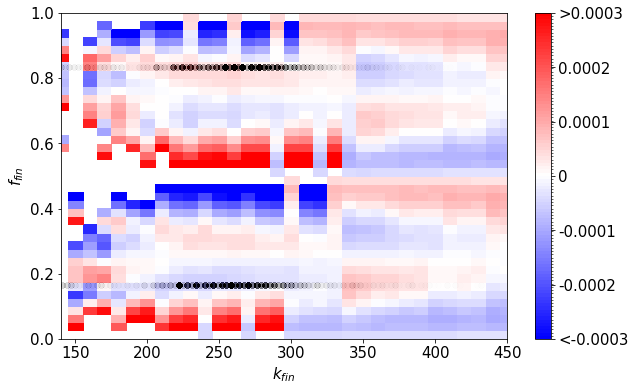} 	\label{fig:heatmap_p025}
    }

\caption{The same realized values of $(k_\mathrm{fin},f_\mathrm{fin})$ as Figure \ref{fig:ffin} are plotted on the heatmap of $\tilde{b}(k_\mathrm{fin}, f_\mathrm{fin})$ calculated by Algorithm \ref{alg:EstimBias}. $\tilde{b}(k_\mathrm{fin}, f_\mathrm{fin})$ indicates the bias generated in the final round of IQAE conditioned on the Grover number $k_\mathrm{fin}$ and $f_\mathrm{fin}$ in Eq. \eqref{eq:f_fin}. Now, the case of $a=0.25$ is added. Again, the points are in transparent black, and thus darker regions indicate that there are many realizations in it and lighter regions indicate the opposite. In the heatmap, the color of the regions with NaN $\tilde{b}(k_\mathrm{fin}, f_\mathrm{fin})$ is set to white. For $a=0.2505$, the realized values of $(k_\mathrm{fin},f_\mathrm{fin})$ are concentrated on a few lines, whereas for $a=0.2006$, $(k_\mathrm{fin},f_\mathrm{fin})$ is distributed more broadly. For $a=0.25$, the distribution is concentrated on a few lines but has a reflection symmetry about $f_{\rm min}=0.5$.}
\label{fig:heatmap}
\end{figure*}

We now plot the distribution of $(k_\mathrm{fin},f_\mathrm{fin})$ on the heatmap of $\tilde{b}(k_\mathrm{fin}, f_\mathrm{fin})$, which is highly illustrative for understanding why the bias is enhanced for specific values of $a$.
In Figure \ref{fig:heatmap}, we show the results for $a=0.2505, 0.2006$, and, in addition, $a=0.25$, for which the 10,000 runs of Algorithm \ref{alg:IQAE}$^\prime$ are conducted too.

We obtain $b(a)$ by averaging the values of $\tilde{b}(k_\mathrm{fin}, f_\mathrm{fin})$ over the realized values of $(k_\mathrm{fin}, f_\mathrm{fin})$, which is represented by the black points, with the weight proportional to the realization frequency, which is represented by the darkness of the points.
In the case of $a=0.2505$, as seen above, the realized values of $(k_\mathrm{fin}, f_\mathrm{fin})$ are not broadly distributed but located on a few specific lines.
Contrary to this, in the case of $a=0.2006$, the points are distributed more broadly, and the values of $\tilde{b}(k_\mathrm{fin}, f_\mathrm{fin})$ on the points take various values, both positive and negative, canceling out each other when averaged.
This cancellation tends not to occur in the case of $a=0.2505$ due to the concentration of the points.
In fact, we see that the lines consisting of the realized values of $(k_\mathrm{fin}, f_\mathrm{fin})$ mainly go through the regions where $\tilde{b}(k_\mathrm{fin}, f_\mathrm{fin})$ is positive, which leads to positive $b(a)$.

Of course, even if Eq. (\ref{eq:LargeBiasTheta}) holds and thus the realized values of $(k_\mathrm{fin}, f_\mathrm{fin})$ concentrate on a few lines, the values of $\tilde{b}(k_\mathrm{fin}, f_\mathrm{fin})$ on the points may accidentally cancel out each other, resulting in a relatively small $b(a)$.
Conversely, even if the points are distributed broadly, the distribution is not uniform and the heatmap of $\tilde{b}(k_\mathrm{fin}, f_\mathrm{fin})$ is not completely antisymmetric, which means that the bias cancellation is not perfect.
Nevertheless, the distribution of realized $(k_\mathrm{fin}, f_\mathrm{fin})$ and the cancellation of $\tilde{b}(k_\mathrm{fin}, f_\mathrm{fin})$ on the points roughly describe the phenomenon that the bias is enhanced for the specific values of $a$.

We remark on the case that $\theta_a$ can be exactly written in the form of $l\pi/2m$ with small $m$.
For example, $a=0.25$ gives $\theta_a=l\pi/2m$ with $l=1$ and $m=3$.
Although $a=0.25$ has only a slight difference from $a=0.2505$, $a=0.25$ does not yield the large bias: for it, $b(a=0.25)=3.5\times10^{-6}$, which is one order of magnitude smaller than $b(a=0.2505)=3.7\times10^{-5}$.
Figure \ref{fig:heatmap_p025} shows the distribution of realized $(k_\mathrm{fin}, f_\mathrm{fin})$ on the heatmap of $\tilde{b}(k_\mathrm{fin}, f_\mathrm{fin})$ for $a=0.25$.
We see that the points are distributed on the two horizontal lines\footnote{Although $f_\mathrm{fin}$ can take $\frac{1}{2}$ for $a=0.25$, none of the 10,000 runs ended with $f_\mathrm{fin}=\frac{1}{2}$, and thus there is no point on the line $f_\mathrm{fin}=\frac{1}{2}$ in Figure \ref{fig:heatmap_p025}.} located at the symmetric positions with respect to reflection about the line $f_\mathrm{fin}=1/2$.
Since $\tilde{b}(k_\mathrm{fin}, f_\mathrm{fin})$ is nearly antisymmetric about this line, its values at the realized points largely cancel out each other, yielding the relatively small $b(a)$.

\subsection{Mitigation of the bias \label{sec:miti}}

Finally, we propose a simple method to mitigate the bias: just re-executing the final round.
Namely, we modify Algorithm \ref{alg:IQAE} as follows.
Suppose that we get the CI of $a$ such that $\Delta a_{i,j} \le \epsilon$ in the round with the Grover number $k_\mathrm{fin}$ and the total shot number $N_\mathrm{fin}$.
Then, we perform just one additional round using the same $k_\mathrm{fin}$ and $N_\mathrm{fin}$, and let the resultant maximum likelihood estimate $\hat{a}$ in the added round be the output of the algorithm.
In this additional round, we do not impose the termination criterion $\Delta a_{i,j} \le \epsilon$.
We make $N_\mathrm{fin}$ measurements on $G^{k_\mathrm{fin}}\ket{\Phi}$ even if $\Delta a_{i,j} \le \epsilon$ is satisfied in the middle.
We show the modified algorithm as Algorithm \ref{alg:IQAERedo}. 

\begin{figure}[H]
\begin{algorithm}[H]
\caption{Modified IQAE with the final round re-executed}
\label{alg:IQAERedo}
\begin{algorithmic}[1]
\Require $\epsilon,\alpha \in (0,1)$, $N_{\text{shot}} \in \mathbb{N}$, $r_\mathrm{min}>1$

\State Run Algorithm \ref{alg:IQAE}. Let the Grover number, the total shot number, and $R_i$ in the final round be $k_\mathrm{fin}$, $N_\mathrm{fin}$, and $R_\mathrm{fin}$, respectively.
\State Iterate generating $G^{k_\mathrm{fin}}\ket{\Phi}$ and measuring it $N_\mathrm{fin}$ times. Let the number of times $\ket{\phi_1}$ is obtained be $N_1$.
\State \Return
\begin{equation}
    \sin^2\left(\frac{R_\mathrm{fin}\frac{\pi}{2}+\gamma\left(\sqrt{\frac{N_1}{N_\mathrm{fin}}},R_\mathrm{fin}\right)}{2k_\mathrm{fin}+1}\right)
\end{equation}
as $\hat{a}$.
\end{algorithmic}
\end{algorithm}
\end{figure}

The reason why we expect that this re-execution-based bias mitigation works is as follows.
Recall that, in Algorithm \ref{alg:IQAE}, the error is generated only in the final round as long as the second-to-last round yields the CI of $a$ enclosing the true value, and the bias is induced by the termination criterion $\Delta a_{i,j} \le \epsilon$.
We thus consider that running an additional round without the criterion $\Delta a_{i,j} \le \epsilon$ and taking the result as the final output will mitigate the bias.

\begin{figure}[tp]
\centering
\subfigure[Absolute average error.]{
    \includegraphics[width=1\columnwidth]{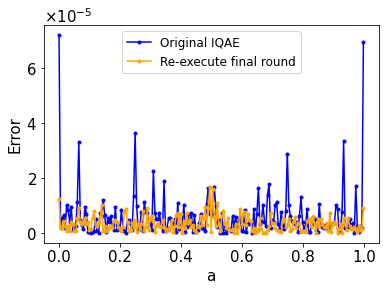} 	\label{fig:BiasRedo}
    }
\subfigure[Average of the total query number.]{
    \includegraphics[width=1\columnwidth]{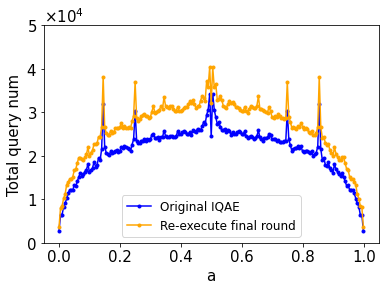} 	\label{fig:QueryNum}
    }
\caption{For the various values of the amplitude $a$, we plot (a) $|\bar{b}(a)|$, the absolute value of the average of the errors in the 10,000 runs of IQAE (Eq. (\ref{eq:avgErr})) and (b) the average number of queries to $G$ in the same runs of IQAE. The blue line denotes the result for Algorithm \ref{alg:IQAE}, which does not incorporate bias mitigation, and the orange one denotes that for Algorithm \ref{alg:IQAERedo}, which includes bias mitigation.}
\end{figure}

In Figure \ref{fig:BiasRedo}, we plot $\left|\bar{b}(a)\right|$ in Algorithm \ref{alg:IQAE}, which is plotted also in Figure \ref{fig:BiasForVariousA}, and that in Algorithm \ref{alg:IQAERedo}.
As this figure shows, the enhanced biases for the specific values of $a$ in Algorithm \ref{alg:IQAE} are largely reduced in Algorithm \ref{alg:IQAERedo}, which indicates that re-executing the final round mitigates the bias.
If we take, among the examined values of $a$, those for which $\left|\bar{b}(a)\right|\ge2\bar{\sigma}(a)$, the average and maximum of the bias reduction rates for these values of $a$ are $57.8\%$ and $99.2\%$, respectively. 

An obvious drawback of this bias mitigation method is the increase of the total number of queries to $G$ by running an additional round.
We have confirmed that, at least in our numerical experiments, this query number increase is mild.
Figure \ref{fig:QueryNum} shows the average number of queries to $G$ in the 10,000 run of Algorithm \ref{alg:IQAE} and Algorithm \ref{alg:IQAERedo} \footnote{The reason why the curves in Figure \ref{fig:QueryNum} have several peaks is as follows. In each round in Algorithm \ref{alg:IQAE}, we search the next Grover number $k_{i+1}$ by Algorithm \ref{alg:find-next-k}, and depending on the current Grover number $k_i$ and the true angle $\theta_a$, this search tends to require many shots in order to narrow down the CI $[\theta^l_{i,j}, \theta^u_{i,j}]$ of $\theta_a$ so much that $[K_{i+1}\theta^l_{i,j}, K_{i+1}\theta^u_{i,j}]$ lies within a single quadrant. This can happen for any $\theta_a$, but for specific values of $\theta_a$, the probability of this phenomenon is relatively high, which leads to the larger total query number in expectation.}.
The ratio of the average query number is about 1.25 for any examined value of $a$.
This mild increase is because the final round in Algorithm \ref{alg:IQAE} is not dominant among all the rounds with respect to query number.
Namely, on average, the final round accounts for about 25\% of the total number of queries to $G$ across all the rounds.

\section{Summary \label{sec:Sum}}

In this paper, we focused on the bias in IQAE, a widely studied version of QAE.
We saw that the bias is enhanced for the specific values of the estimated amplitude $a$.
The termination criterion that the estimated accuracy $\Delta a_{i,j}$ of $a$ falls below the predetermined accuracy $\epsilon$ is a source of the bias.
Decomposing the bias into the bias conditioned by $k_\mathrm{fin}$ the Grover number in the final round, we found a key factor that determines the magnitude of the bias: the distribution of the realized values of $(k_\mathrm{fin},f_\mathrm{fin})$ in the landscape of the conditional bias.
Here, $f_\mathrm{fin}$ is defined as Eq. (\ref{eq:f_fin}), and a main factor to determine the probability distribution of the IQAE estimate $\hat{a}$ and thus its bias.
We found that for $a$ such that Eq. (\ref{eq:LargeBiasTheta}) holds with small $m$ and tiny $\delta$, the points of realized $(k_\mathrm{fin},f_\mathrm{fin})$ are located only on a few lines in a 2D plane, and the conditional biases at the points do not tend to cancel each other so much, resulting in the large bias.
We also proposed a simple bias mitigation method by re-executing the final round with the same Grover number $k_\mathrm{fin}$ and shot number $N_\mathrm{fin}$.
We saw that the increase of the total number of queries to $G$ is mild, about 25\%.

\section*{Acknowledgements}

Not applicable.

\section*{Abbreviations}

QAE, quantum amplitude estimation; IQAE, iterative quantum amplitude estimation; QPE, quantum phase estimation; QMCI, quantum algorithm for Monte Carlo integration; MLE, maximum likelihood estimation; MLEQAE, maximum likelihood estimation-based quantum amplitude estimation; CI, confidence interval;

\section*{Declaration}

\subsection*{Ethical approval and consent to participate}

Not applicable.

\subsection*{Consent for publication}

The author has approved the publication. The research in this work did not involve any human, animal or other
participants.

\subsection*{Availability of supporting data}

The data that support the findings of this study can be obtained from the corresponding author upon a reasonable
request.

\subsection*{Competing interests}

The author declares no competing interests.

\subsection*{Authors' contributions}

KM as the sole author of the manuscript, conceived, designed, and performed the analysis; he also wrote and reviewed the paper. The author read and approved the final manuscript.

\subsection*{Funding}

The author is supported by MEXT Quantum Leap Flagship Program (MEXT Q-LEAP) Grant no. JPMXS0120319794, JSPS KAKENHI Grant no. JP22K11924, and JST COI-NEXT Program Grant No. JPMJPF2014.

\bibliography{reference}

\end{document}